%% file: main.tex
\documentclass[journal]{IEEEtran}
\usepackage{subcaption}
\usepackage{amsmath, amssymb, bm, mathrsfs, xfrac, mathtools, amsthm}
\usepackage{lettrine}
\usepackage{pgf}
\usepackage{booktabs}
\usepackage{amsmath, amssymb, bm, xfrac, mathtools, amsthm}
\usepackage{circuitikz}
\usepackage{mathtools} %
\usepackage{empheq}
\usepackage[export]{adjustbox}
\usepackage{cuted} 
\usepackage{capt-of}
\usepackage{cite}
\usepackage{afterpage}
\usepackage{dblfloatfix}
\usepackage[table]{xcolor} 
\usepackage{hyperref}
\usepackage{url}
\usepackage{standalone}
\usepackage{tikz}\usetikzlibrary{calc, shapes, arrows, automata, positioning, patterns}
\ifCLASSINFOpdf
\else
\fi

\newcommand{\gfl}{q_{\mathrm{gfl}}}
\newcommand{\gfm}{q_{\mathrm{gfm}}}
\newcommand{\ts}{t_s}

\begin{document}
\bstctlcite{IEEEexample:BSTcontrol}
\title{Dynamic state estimation of hybrid systems: Inverters that switch between grid-following and grid-forming control schemes
{\small }} 
\author{Bukunmi G. Odunlami,~\IEEEmembership{Student Member,~IEEE}, Marcos Netto,~\IEEEmembership{Senior Member,~IEEE}
\thanks{This work is supported by the National Science Foundation under Grant 2328241. The authors are with the Department of Electrical and Computer Engineering, New Jersey Institute of Technology, Newark, NJ 07102, USA.}
}
\maketitle

\begin{abstract}
This paper develops a hybrid system modeling framework for inverters that switch between grid-following and grid-forming control schemes. In particular, such inverters are modeled as hybrid automata with guard conditions on voltage and frequency, and reset maps that maintain consistent phase, frequency, and droop references during mode transitions. The hybrid model is embedded within an extended Kalman filter to assess estimation performance under explicit mode switching. Results show that the proposed framework ensures stable, well-behaved dynamics and improves state estimation, especially near switching instants, compared with smooth continuous models.
\end{abstract}


\IEEEpeerreviewmaketitle

\section{Introduction}
\emph{Hybrid dynamics}---emerging from systems that exhibit both continuous and discrete behaviors are ubiquitous in power grids. Examples include dynamics involving switching events triggered by on-load tap changers, protection relays, and load shedding schemes \cite{Hiskens2000, Susuki2012}. 

Still, continuous-time ordinary differential-algebraic models are widely used to analyze the dynamic performance of power grids \cite{Sauer1997}. Generators are modeled by differential equations whose structure and parameters remain constant over time, and the network interconnecting them is represented by algebraic equations. Hence, switching events such as those mentioned earlier are conveniently represented by adjusting admittance values in the algebraic equations. Analyses are then separated into pre-fault, fault-on, and post-fault network conditions.

However, the dynamics of inverter-based resources (IBRs) can introduce structural changes to the differential equations describing generators \cite{Odunlami2025a}. For example, the overcurrent limiter mechanism, when activated, modifies the inverter's control scheme. Once engaged, the current-control loop saturates, and the terminal voltage follows an alternative regulation path \cite{10603443}. Another example is switching reactive power control modes based on grid operating conditions, commonly used by utility-scale photovoltaic plants \cite{Mishra2026}. 

Indeed, the rise of IBRs exacerbates hybrid dynamics and prompts questions about the parsimony of continuous-time models that approximate switching by interpolating between control laws, enforcing artificial smoothness on fundamentally discontinuous dynamics \cite{Badrzadeh2024b}. This mismatch can lead to misleading conclusions about the system dynamics and stability \cite{GOEBEL20041}. As IBR control schemes rapidly evolve, these challenges are becoming increasingly critical.

First-generation IBRs draw power into the grid at a unity power factor and seldom provide grid services, such as voltage regulation \cite{Chaudhuri2024}. These IBRs, known as grid-following inverters (GFLs), synchronize with the grid via a phase lock loop (PLL). GFLs are reliable and commonly employed, but they require a minimum system strength e.g., a weighted short-circuit ratio of at least 1.5 \cite{Badrzadeh2021} for the PLL to synchronize and maintain synchronism with the grid; otherwise, GFLs can compromise system stability. Conversely, grid-forming inverters (GFMs) synchronize with the grid using droop control and can help regulate voltage and frequency. GFMs are now becoming a reality \cite{Badrzadeh2024}; they can provide grid services, thereby offering essential support during faults and even black-start conditions.

Although GFLs and GFMs share structural similarities and can even be considered duals, they excel under different grid conditions \cite{9714816}. GFLs function well when the system is strong, but may lose synchronism otherwise. On the other hand, GFMs can enhance stability when the system is weak, but may cause unwanted oscillations when the system is strong \cite{5456209}. Indeed, regulatory agencies \cite{NERC2021} recognize this trade-off and require IBRs to remain connected to provide support during faults. It is possible to combine the complementary characteristics of GFLs and GFMs into a single control system \cite{Geng2022}, or to switch between control modes depending on the system's operating conditions. For example, \cite{seamless} describes a switching method that manages the different outer control loops of GFLs and GFMs and ensures control setpoints are adjusted to maintain consistent operating points during transitions. This paper develops a hybrid system formulation for such inverters that switch between GFL and GFM control schemes.

We show that explicitly representing hybrid dynamics is both feasible and of paramount importance for dynamic state estimation, particularly in a decentralized setting \cite{Tan2023}. To assess the analytical and estimation implications of this comprehensive modeling approach, the hybrid system is embedded within an extended Kalman filter (EKF) \cite{Netto2016} that propagates uncertainty through each mode transition using the saltation matrix \cite{KONG2021109752}. This paper is the first to propose a hybrid system formulation for dynamic state estimation and proceeds as follows. Section II briefly introduces the continuous-time model of GFLs and GFMs. Section III develops the hybrid system formulation and derives an EKF-based, dynamic state estimator for hybrid systems. Section IV presents numerical results, and Section V concludes the paper.

\section{Preliminaries}
\begin{figure}[!t]
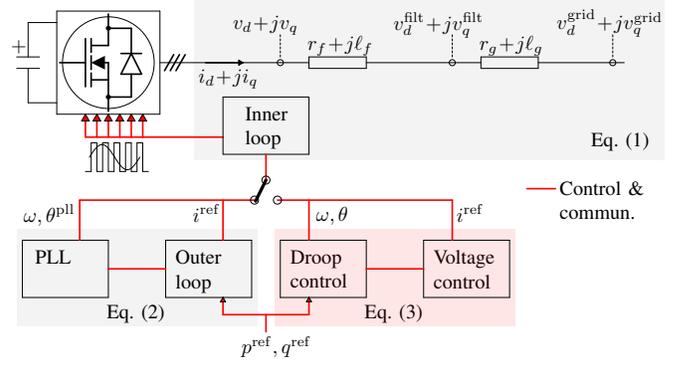
\centering
\includestandalone[width=1\columnwidth]{fig1}
\vspace{-.6cm}
\caption{Inverter switching between GFL and GFM controls.}
\label{fig:switchingcontrols}
\end{figure}

The modeling in this section follows the continuous-time approach commonly used in the literature \cite{Venkatramanan2022}. Consider an inverter that switches between GFL and GFM control modes (Fig. \ref{fig:switchingcontrols}) depending on grid operating conditions. We start by noting that the modeling of the filter and the inner loop is the same for GFLs and GFMs, as given by:
\begin{equation}
\mathbf{f}_{\mathrm{common}}\!:\!
\begin{cases}
\begin{rcases}
\dot{\gamma}_d = k_i^{c}\,(i_d^{\mathrm{ref}} - i_d) \\
\dot{\gamma}_q = k_i^{c}\,(i_q^{\mathrm{ref}} - i_q) \\
v_d = k_p^{c}\,(i_d^{\mathrm{ref}} - i_d) + \gamma_d + \omega\,\ell_f\, i_q \\
v_q = k_p^{c}\,(i_q^{\mathrm{ref}} - i_q) + \gamma_q - \omega\,\ell_f\, i_d
\end{rcases}\text{\shortstack{Inner\\loop}}\\
\begin{rcases}
\left({\ell_g}/{\omega_b}\right)\dot{i}_d = v_d^{\mathrm{filt}} - v_d^{\mathrm{grid}} - r_g\, i_d + \omega\,\ell_g\, i_q \\
\left({\ell_g}/{\omega_b}\right)\dot{i}_q = v_q^{\mathrm{filt}} - v_q^{\mathrm{grid}} - r_g\, i_q - \omega\,\ell_g\, i_d
\end{rcases}\text{Filter}
\end{cases}
\label{eq:common_dynamics}
\end{equation}

\noindent
where $\vartheta_d$ ($\vartheta_q$) indicates the direct (quadrature) component of a variable $\vartheta$; $i_d^{\mathrm{ref}}$ and $i_q^{\mathrm{ref}}$ are setpoints from outer controllers; $k_p^{c}$ ($k_i^{c}$) is the proportional (integral) gain of a PI controller in the inner loop; $\omega_b$ is the base frequency; and the instantaneous frequency $\omega$ equals $\dot\theta$ ($\dot\theta^{\mathrm{pll}}$) when the inverter operates as GFM (GFL). The other variables are explicit in Fig. \ref{fig:switchingcontrols}.

When the inverter operates as a GFL, a synchronous reference frame PLL estimates the grid angle, $\theta^{\mathrm{pll}}$, and frequency, $\omega=\dot\theta^{\mathrm{pll}}$, thereby defining the $dq$ frame. Using synchronized quantities, instantaneous real and reactive power $(p,q)$ are low-pass filtered to $(p_m,q_m)$ and compared with setpoints $(p^{\mathrm{ref}},q^{\mathrm{ref}})$ to generate current commands $(i_{d}^{\mathrm{ref}},i_{q}^{\mathrm{ref}})$. Formally,
\begin{equation}
\mathbf{f}_{\mathrm{gfl}}:
\begin{cases}
\mathbf{f}_{\mathrm{common}} \\
\begin{rcases}
\dot{\eta}^{\mathrm{pll}} = k_i^{\mathrm{pll}}\, v_q^{\mathrm{filt}}(\theta^{\mathrm{pll}}) \\
\dot\theta^{\mathrm{pll}} = \omega = \omega_0 + k_p^{\mathrm{pll}}\, v_q^{\mathrm{filt}}(\theta^{\mathrm{pll}}) + \eta^{\mathrm{pll}}
\end{rcases}\text{PLL}\\
\begin{rcases}
\dot{\sigma}_p = p^{\mathrm{ref}} - p_m \\
\dot{\sigma}_q = q^{\mathrm{ref}} - q_m \\
i_d^{\mathrm{ref}} = k_p^{p}\,(q^{\mathrm{ref}} - q_m) + k_i^{p}\,\sigma_q \\
i_q^{\mathrm{ref}} = k_p^{q}\,(p^{\mathrm{ref}} - p_m) + k_i^{q}\,\sigma_p
\end{rcases}\text{\shortstack{Outer\\loop}}\\
\end{cases}
\label{eq:fgfl}
\end{equation}

\noindent
where $k_p^{\mathrm{pll}}$ and $k_p^{p}$ ($k_i^{\mathrm{pll}}$ and $k_i^{p}$) are the proportional (integral) gains of a PI controller in the PLL and outer loop, respectively.

\noindent
The state vector $\mathbf{x}_{\mathrm{gfl}} = [\, \gamma_d\; \gamma_q\; i_d\; i_q\; \eta^{\mathrm{pll}}\; \theta^{\mathrm{pll}}\; \sigma_p\; \sigma_q\,]^{\top}$, and the input vector $\mathbf{u}_{\mathrm{gfl}} = [\,p^{\mathrm{ref}}\; q^{\mathrm{ref}} \,]^{\top}$.

When the inverter operates as a GFM, droop control eliminates the need for explicit synchronization, such as with a PLL, because it provides an inherent synchronization mechanism. It computes $\omega$ and $v^{\mathrm{ref}}$ from $(p_m,q_m)$ and defines the reference angle~$\theta$. The voltage control loop regulates~$v^{\mathrm{ref}}$ to yield current 
commands $(i_d^{\mathrm{ref}},i_q^{\mathrm{ref}})$. Formally,
\begin{equation}
\mathbf{f}_{\mathrm{gfm}}:
\begin{cases}
\mathbf{f}_{\mathrm{common}} \\
\begin{rcases}
\dot{\theta} = \omega = \omega_0 - m_p\,(p_m - p_0) \\
v^{\mathrm{ref}} = v_0 - n_q\,(q_m - q_0)
\end{rcases}\text{\shortstack{Droop\\control}}\\
\begin{rcases}
\dot{\xi}_d = k_i^{v}\,(v^{\mathrm{ref}} - v_d) \\
\dot{\xi}_q = k_i^{v}\,(0 - v_q) \\
i_d^{\mathrm{ref}} = k_p^{v}\,(v^{\mathrm{ref}} - v_d) + \xi_d \\
i_q^{\mathrm{ref}} = k_p^{v}\,(0 - v_q) + \xi_q
\end{rcases}\text{\shortstack{Voltage\\control}}\\
\end{cases}
\label{eq:fgfm}
\end{equation}

\noindent
where $k_p^{v}$ ($k_i^{v}$) denotes the proportional (integral) gain of the PI controller in the voltage control loop, and $m_p$ and $n_p$ are droop control coefficients. The state vector $\mathbf{x}_{\mathrm{gfm}} = [\,\gamma_d\; \gamma_q\; i_d\; i_q\; \theta\; \xi_d$ $\xi_q \,]^{\top}$, and the input vector $\mathbf{u}_{\mathrm{gfm}} = [\,p^{\mathrm{ref}}\; q^{\mathrm{ref}} \,]^{\top}$. Note that \eqref{eq:fgfl} and \eqref{eq:fgfm} are different in structure, and $\mathrm{dim}(\mathbf{x}_{\mathrm{gfl}})\neq\mathrm{dim}(\mathbf{x}_{\mathrm{gfm}})$.

Next, using \eqref{eq:common_dynamics}--\eqref{eq:fgfm}, we develop a hybrid dynamical system formulation for the inverter in Fig. \ref{fig:switchingcontrols}.

\section{Hybrid system and dynamic state estimation}
Let a hybrid automaton $\mathcal{H} \coloneq \langle \mathrm{Q}, \mathrm{X}, \mathrm{F}, \mathrm{Init}, \mathrm{Inv}, \mathrm{G}, \mathrm{R} \rangle$. The set $\mathrm{Q}$ contains the inverter's discrete modes of operation. The states in \eqref{eq:fgfl}--\eqref{eq:fgfm} are in the set $\mathrm{X} \subseteq \mathbb{R}^n$. The flow set $\mathrm{F}$ defines the continuous dynamics associated with each discrete mode $q \in \mathrm{Q}$. The initial set $\mathrm{Init} \subseteq \mathrm{Q} \times \mathrm{X}$ specifies the admissible initial conditions. The invariant set $\mathrm{Inv}(q)$ defines the space where the system's continuous state is allowed to evolve while it stays in a mode $q$. The set $\mathrm{G}$ defines the guard conditions for mode transition. Finally, the reset map $\mathrm{R}$ specifies how the system states and control parameters are updated at the switching instant. Formally,
\vspace{-.2cm}\begin{align}
\mathcal{H} \coloneqq
\begin{cases}
\mathrm{Q} = \{\gfl, \gfm\} \\
\mathrm{X} = \{\mathbf{x}_\mathrm{gfl}, \mathbf{x}_\mathrm{gfm}\} \\
\mathrm{F} =
\begin{cases}
\dot{\mathbf{x}}_{\mathrm{gfl}} = \mathbf{f}_{\mathrm{gfl}}(\mathbf{x}_{\mathrm{gfl}},\mathbf{u}_{\mathrm{gfl}}) & \text{if}\;q = \gfl \\
\dot{\mathbf{x}}_{\mathrm{gfm}} = \mathbf{f}_{\mathrm{gfm}}(\mathbf{x}_{\mathrm{gfm}},\mathbf{u}_{\mathrm{gfm}}) & \text{if}\;q = \gfm
\end{cases} \\
\mathrm{Init} = \{\gfl\} \times \{\mathbf{x}_{\mathrm{gfl}}\} \\
\mathrm{Inv}(q_{\mathrm{gfl}}) = \{\, |v^{\mathrm{grid}}| \in v_{\mathrm{th}} \;\wedge\; |\Delta \omega| \in \omega_{\mathrm{th}} \,\} \\[4pt]
\mathrm{G}(q_{\mathrm{gfl}}, q_{\mathrm{gfm}}) = \{\, |v^{\mathrm{grid}}| \notin v_{\mathrm{th}} \;\vee\; |\Delta \omega| \notin \omega_{\mathrm{th}} \,\}\\
\mathrm{R}(q_1, q_2, \mathbf{x}_\mathrm{gfl}) \;=\;
\{\, \mathbf{x}_\mathrm{gfm} :
\theta_\mathrm{gfm}^{+} = \theta^{-},\;
\omega_\mathrm{gfm}^{+} = \omega^{-},\\
\;p_{0} = p_{0}^{+}, 
q_{0} = q_{0}^{+}, \; r^+, \; \ell^+ \}\\
\mathrm{R}(q_2, q_1, \mathbf{x}_\mathrm{gfm}) \;=\;
\{\, \mathbf{x}_\mathrm{gfl} :
\theta_\mathrm{gfl}^{+} = \theta^{-},\;
\omega_\mathrm{gfl}^{+} = \omega^{-} \}
\end{cases} \label{HA}
\end{align}

\vspace{-.3cm}
\noindent
where $\mathcal{A}\times\mathcal{B}$ indicates the cross product of the elements in the sets $\mathcal{A}$ and $\mathcal{B}$; $v_\mathrm{th}$ and $\omega_\mathrm{th}$ are predefined thresholds; $\wedge$ (resp., $\vee$) denotes the \emph{logical and} (resp., \emph{logical or}), and $\cdot^{-}$ ($\cdot^{+}$) denotes a quantity immediately before (after) the switch---or \emph{jump}, in hybrid system terminology. For instance, consider a representative operating scenario in which the inverter in Fig. \ref{fig:switchingcontrols} switches from GFL to GFM when grid conditions depart from normal operating ranges. Specifically, the switch occurs when the grid voltage magnitude $|v_\mathrm{grid}|$ falls outside an admissible band, $v_\mathrm{th}$, or the frequency deviation $|\Delta \omega|$ exceeds a tolerance range, $\omega_\mathrm{th}$. Such events typically arise during low-voltage ride-through or weak-grid conditions, prompting the inverter to enter GFM mode to support voltage and frequency. Just before the switch instant, $t_s$, we sample
\vspace{-.15cm}
\begin{align}
\theta^- &= \theta^{\mathrm{pll}}(\ts^-) &
p_s &= v_d^- i_d^- + v_q^- i_q^-\nonumber \\
\omega^- &= \omega^{\mathrm{pll}}(\ts^-) &
q_s &= v_q^- i_d^- - v_d^- i_q^- 
\end{align} 

\vspace{-.15cm}
\noindent
Furthermore, we design the reset maps in \eqref{HA} to capture the interaction between the GFL and GFM control schemes and to maintain consistent dynamic behavior at the switching instant. Specifically, for the transition from $q_{\mathrm{gfl}}$ to $q_{\mathrm{gfm}}$, we consider: 

\emph{$P$–$\omega$ droop bias:} The PLL-derived estimates of frequency and phase are transferred at the switching instant, such that $\omega_{\mathrm{gfm}}^{+} = \omega^{-}$ and $\theta_{\mathrm{gfm}}^{+} = \theta^{-}$. The corresponding droop bias is then recomputed and substituted into \eqref{eq:fgfm} for $t = t_s^+$,
\vspace{-.15cm}
\begin{align}
p^{+}_{0} = p_s - \frac{\omega_0 - \omega^-}{m_p}
\end{align}

\vspace{-.15cm}
\emph{Q–V droop bias:} For a bumpless transfer, $v^\text{ref}$ is set equal to the grid voltage magnitude immediately before the switching, $|v^{\mathrm{grid}}|^-$. The reactive power droop bias is then recomputed and substituted into \eqref{eq:fgfm} for $t = t_s^+$,
\vspace{-.15cm}
\begin{align}
q^{+}_{0} = q_s - \frac{v_0 - |v^{\mathrm{grid}}|^-}{n_q}
\end{align}

\vspace{-.15cm}
\emph{Current-limit through threshold virtual impedance:} The instantaneous output current magnitude $i_s$ is evaluated. Following \cite{10603443}, a linear activation function $\psi$ is defined to scale the virtual impedance between the threshold $i_{\mathrm{th}}$ and the maximum current $i_{\max}$. The effective filter parameters are then updated as $r^{+} = (r_f + r_g) \psi r_{\mathrm{vi}}$ and $\ell^{+} = (\ell_f + \ell_g)\psi \ell_{\mathrm{vi}}$, where
\vspace{-.2cm}
\begin{equation}
\psi =
\begin{cases}
0 & i_s \le i_{\mathrm{th}} \\
\dfrac{i_s - i_{\mathrm{th}}}{i_{\max} - i_{\mathrm{th}}} & i_{\mathrm{th}} < i_s < i_{\max} \\
1 & i_s \ge i_{\max}
\end{cases}
\end{equation}

\vspace{-.1cm}
\noindent
$i_s = \sqrt{(i_d^-)^2 + (i_q^-)^2}$, and $r_{\mathrm{vi}}$ ($\ell_{\mathrm{vi}}$) denotes virtual resistance (inductance).

On the other hand, the reverse transition from $q_{\mathrm{gfm}}$ to $q_{\mathrm{gfl}}$ is triggered when the grid voltage magnitude remains within the admissible band $v_{\mathrm{th}}$ for a dwell time exceeding a prescribed threshold $T_{\mathrm{hold}}$, ensuring that the system does not immediately revert to GFL following brief voltage recoveries. Accordingly, the associated reset map is reduced to the identity map $\mathrm{R}(q_{\mathrm{gfm}}, q_{\mathrm{gfl}}, \mathrm{X})=\mathrm{X}$ and the phase-angle continuity is preserved across the switching instant, such that $\theta_{\mathrm{gfl}}^{+} = \theta_{\mathrm{gfm}}^{-}$ and $\omega_{\mathrm{gfl}}^{+} = \omega_{\mathrm{gfm}}^{-}$, ensuring a smooth re-entry of the GFL dynamics without introducing spurious jumps.

We are now in the position to derive a dynamic state estimator for the hybrid system \eqref{HA}.

\begin{figure}[!t]
\centering
\begin{tikzpicture}[->, >=stealth', auto, semithick, node distance=4cm, scale=0.8, every node/.style={scale=0.8}]
\node[state, minimum size=1.5cm, align=center] (q1) {\text{GFL} \\ $\mathbf{f}(q_1, \mathbf{x}, \mathbf{u})$ \\ $\mathrm{G}(q_1, q_2)$};
\node[state, minimum size=1.2cm, align=center, right of=q1] (q2) {\text{GFM} \\ $\mathbf{f}(q_2, \mathbf{x}, \mathbf{u})$ \\ $\mathrm{G}(q_2, q_1)$};
\path (q1) edge [bend left] node {$\mathrm{R}(q_1, q_2, \mathrm{X})$} (q2)
(q2) edge [bend left] node {$\mathrm{R}(q_2, q_1, \mathrm{X})$} (q1);
\draw[->, thick, shorten >=2pt] ($(q1.west)+(-1.2,0)$) -- (q1.west)
node[pos=0.3, above] {\large $\mathbf{u}_\text{gfl}$};
\draw[->, thick, shorten >=2pt] ($(q2.east)+(1.2,0)$) -- (q2.east)
node[pos=0.45, above] {\large $\mathbf{u}_\text{gfm}$};
\end{tikzpicture}
\caption{Hybrid automaton model.}
\label{fig:hybrid_automata}
\end{figure}
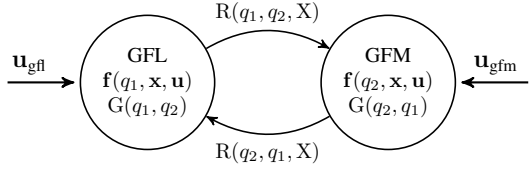

\begin{figure*}[hb!]
\centering
\begin{subfigure}[b]{0.32\textwidth}
\centering
\includegraphics[width=\textwidth]{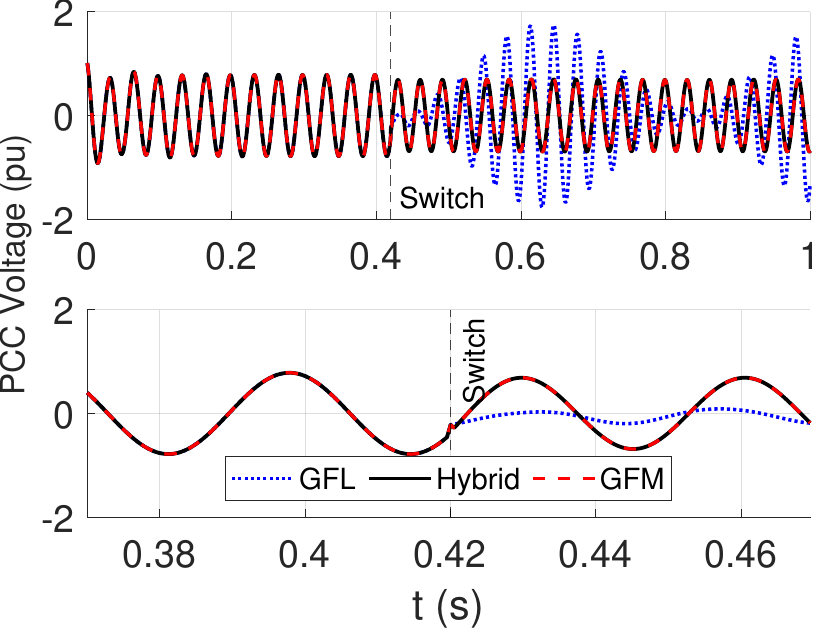}
\caption{Hybrid voltage}
\label{fig:hybrid_voltage}
\end{subfigure}
\hfill
\begin{subfigure}[b]{0.32\textwidth}
\centering
\includegraphics[width=\textwidth]{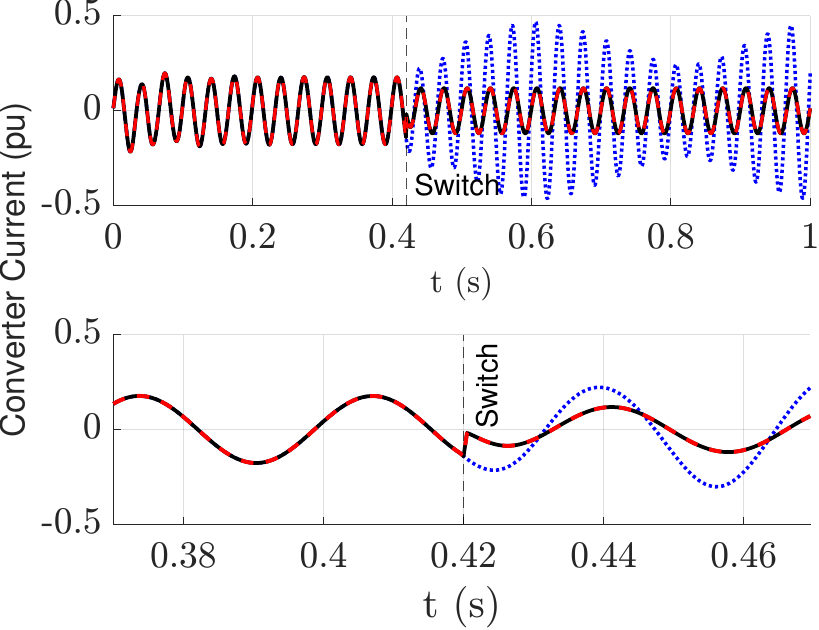}
\caption{Hybrid current}
\label{fig:hybrid_current}
\end{subfigure}
\hfill
\begin{subfigure}[b]{0.32\textwidth}
\centering
\includegraphics[width=\textwidth]{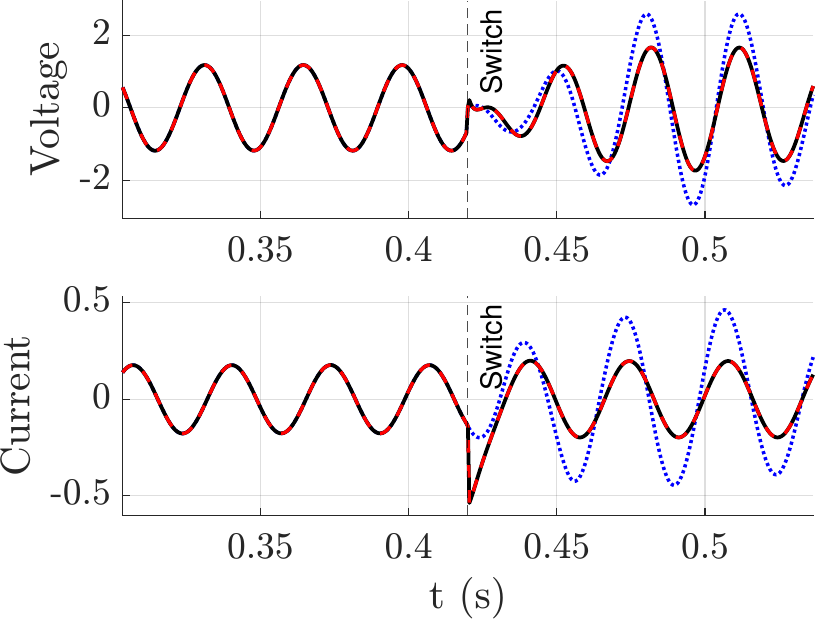}
\caption{Without reset maps}
\label{fig:no_reset}
\end{subfigure}
\caption{Comparison of hybrid model formulations for GFM and GFL modes with and without reset maps.}
\label{fig:hybrid_three_comparison}
\end{figure*}

\subsection{Hybrid extended Kalman filter (EKF)}
Having established the hybrid system representation for the inverter in Fig. \ref{fig:switchingcontrols}, we now derive an EKF that is consistent with \eqref{HA}, where the recursion follows with mode-dependent dynamics and transitions ${q}_{k-1} \!\to\! {q}_k$. At each \emph{prediction step},
\vspace{-.1cm}\begin{align}
\mathbf{x}_{k|k-1} &= \mathbf{f}_{q_{k-1}}(\mathbf{x}_{k-1|k-1},\mathbf{u}_{k-1}) \nonumber \\[-2pt]
\mathbf{P}_{k|k-1} &= \mathbf{F}_{k-1} \mathbf{P}_{k-1|k-1} \mathbf{F}_{k-1}^\top + \mathbf{Q}_{q_{k-1}}
\label{eq:prediction}
\end{align}

\vspace{-.15cm}
\noindent
where a subscript $k$ (resp., $k-1$) indicates the current (resp., previous) prediction step; subscript $q$ indicates dependency on a discrete mode $q$; $\mathbf{f}_{q_{k-1}}$, therefore, denotes mode-dependent, vector-valued, nonlinear functions at step $k-1$; $\mathbf{F}$ denotes the state transition Jacobian; and $\mathbf{P}$ is the prediction covariance. The prediction errors, $\bm{w}_{q_{k-1}}$, are assumed to be independent and identically distributed random processes following a normal distribution, with $\mathbb{E}[\bm{w}_{q_{k-1}}\bm{w}_{q_{k-1}}^\top]=\mathbf{Q}_{q_{k-1}}$.

If $\mathbf{x}_{k|k-1}$ satisfies a guard condition $g(\mathbf{x},\mathbf{u}) = 0$, a discrete transition from mode $q_{k-1}$ to $q_k$ is triggered. At this transition, the state estimate is updated according to a reset map, i.e.,
\vspace{-.15cm}
\begin{equation}
\mathbf{x}_{k|k-1}^{+} \leftarrow \mathrm{R}_{(q_{k-1},q_k)}\!\left(\mathbf{x}_{k|k-1}^{-}\right). \label{priori_guards1}
\end{equation}

\vspace{-.15cm}
To analyze how uncertainty evolves through a switching surface, the notion of a linearized perturbation is introduced. Let $\bar{\mathbf{x}}_{k|k-1}$ denote the nominal predicted trajectory, and
\vspace{-.15cm}
\begin{align}
\mathbf{x}_{k|k-1} = \bar{\mathbf{x}}_{k|k-1} + \delta\mathbf{x}_{k|k-1}
\end{align}

\vspace{-.15cm}
\noindent
where $\delta\mathbf{x}_{k|k-1}$ represents an infinitesimal variation around the nominal trajectory and captures how perturbations in the state propagate through both the continuous and discrete dynamics. This forms the basis for deriving the post-jump covariance. At a reset event, the perturbation evolves according to a saltation matrix, $\mathbf{\Xi}_k$, which governs how deviations are mapped across a mode transition and is derived in the next section. Formally,
\vspace{-.3cm}
\begin{equation}
\delta \mathbf{x}_{k|k-1}^{+}
= \mathbf{\Xi}_k\,\delta \mathbf{x}_{k|k-1}^{-}
+ \mathbf{w}_{\mathrm{R}}
\label{eq:lin_reset}
\end{equation}

\vspace{-.15cm}
\noindent
where the reset error vector, $\mathbf{w}_{\mathrm{R}}$, is assumed zero-mean and independent of $\delta\mathbf{x}_{k|k-1}^{-}$. By definition, the covariances before and after the switching are given by:
\vspace{-.15cm}
\begin{align}
\mathbf{P}_{k|k-1}^{-} = \mathbb{E}\!\big[\delta \mathbf{x}_{k|k-1}^{-}(\delta \mathbf{x}_{k|k-1}^{-})^{\top}\big] \nonumber\\
\qquad 
\mathbf{P}_{k|k-1}^{+} = \mathbb{E}\!\big[\delta \mathbf{x}_{k|k-1}^{+}(\delta \mathbf{x}_{k|k-1}^{+})^{\top}\big] \label{postcovariance}
\end{align}

\vspace{-.15cm}
\noindent 
Substituting \eqref{eq:lin_reset} in \eqref{postcovariance} follows:
\vspace{-.15cm}
\begin{align}
\mathbf{P}_{k|k-1}^{+} 
&= \mathbb{E}\!\Big[(\mathbf{\Xi}_k\delta \mathbf{x}_{k|k-1}^{-} + \mathbf{w}_{\mathrm{R}})
 (\mathbf{\Xi}_k\delta \mathbf{x}_{k|k-1}^{-} + \mathbf{w}_{\mathrm{R}})^{\!\top}\Big] \notag\\
&= \mathbf{\Xi}_k\,\mathbb{E}\!\big[\delta \mathbf{x}_{k|k-1}^{-}(\delta \mathbf{x}_{k|k-1}^{-})^{\!\top}\big]\mathbf{\Xi}_k^{\top}
 + \mathbb{E}\!\big[\mathbf{w}_{\mathrm{R}}\mathbf{w}_{\mathrm{R}}^{\top}\big] \notag\\
&= \mathbf{\Xi}_k\,\mathbf{P}_{k|k-1}^{-}\,\mathbf{\Xi}_k^{\top} + \mathbf{W}_{\mathrm{R}}
\label{eq:jump_cov_derivation}
\end{align}

\vspace{-.15cm}
\noindent
Given that $\mathbb{E}[\delta \mathbf{x}_{k|k-1}^{-}\mathbf{w}_{\mathrm{R}}^{\top}] = 0$ and $\mathbb{E}\big[\mathbf{w}_{\mathrm{R}}\mathbf{w}_{\mathrm{R}}^{\top}\big] = \mathbf{W}_{\mathrm{R}}$, the covariance update across the reset is
\vspace{-.15cm}
\begin{align}
\mathbf{P}_{k|k-1} &\leftarrow \mathbf{\Xi}_k\,\mathbf{P}_{k|k-1}\,\mathbf{\Xi}_k^{\top} + \mathbf{W}_{\mathrm{R}}.
\label{priori_guards2}
\end{align}
\vspace{-.15cm}
At each \emph{filtering (or correction) step},
\begin{align}
\label{eq:measurementupdate}
\mathbf{K}_k &= \mathbf{P}_{k|k-1} \mathbf{H}_{q_{k}}^\top \big(\mathbf{H}_{q_{k}} \mathbf{P}_{k|k-1} \mathbf{H}_{q_{k}}^\top + \mathbf{\Sigma}_{{q}_k}\big)^{-1} \nonumber \\
\mathbf{x}_{k|k} &= \mathbf{x}_{k|k-1} + \mathbf{K}_k \big(\mathbf{z}_k - \mathbf{h}_{{q}_k}(\mathbf{x}_{k|k-1})\big) \\
\mathbf{P}_{k|k} &= (\mathbf{I} - \mathbf{K}_k \mathbf{H}_{q_{k}}) \mathbf{P}_{k|k-1} (\mathbf{I} - \mathbf{K}_k \mathbf{H}_{q_{k}})^\top + \mathbf{K}_k \mathbf{\Sigma}_{{q}_k} \mathbf{K}_k^\top \nonumber
\end{align}
where $\mathbf{h}_{{q}_k}$ denotes a vector-valued, nonlinear measurement function with Jacobian $\mathbf{H}_{q_{k}}$; $\mathbf{K}_k$ is the Kalman gain; $\mathbf{z}_k$, a vector of sampled measurements, and $\boldsymbol{\Sigma}_{q}$ is the measurement noise covariance matrix. As in \eqref{priori_guards1}--\eqref{priori_guards2}, if a correction step triggers a guard condition, the posteriori state estimate $\mathbf{x}_{k|k}$ (resp., error covariance matrix $\mathbf{P}_{k|k}$) is updated with $\mathrm{R}(\cdot)$ (resp., $\mathbf{\Xi}$).

\subsection{The saltation matrix and guard conditions}
Following \cite{KONG2021109752}, the saltation matrix
\vspace{-.15cm}
\begin{align}
\mathbf{\Xi}_{q_{1} q_{2}}
= \mathbf{D}_{\mathbf{x}} \mathrm{R}_{q_{1} q_{2}}^{-} +
\frac{
\big(\mathbf{f}_{q_{2}}^{+} - \mathbf{D}_{\mathbf{x}} \mathrm{R}_{q_{1} q_{2}}^{-} \mathbf{f}_{q_{1}}^{-}\big)
\left(\nabla \mathbf{g}_{q_{1} q_{2}}^{-}\right)^{\top}
}{
\left(\nabla \mathbf{g}_{q_{1} q_{2}}^{-}\right)^{\top} \mathbf{f}_{q_{1}}^{-}
}
\label{eq:saltation}
\end{align}

\vspace{-.15cm}
\noindent
where $\mathbf{D}_{\mathbf{x}} \mathrm{R_{q_{1} q_{2}}^{-}}$ denotes the Jacobian of the reset map from mode $q_1$ to mode $q_2$; $\mathbf{f}_{q_{1}}^{-}$ ($\mathbf{f}_{q_{2}}^{+}$) denotes the flow map just before (after) the jump, and $\mathbf{g}_{q_{1} q_{2}}^{-}$ are the guard functions from mode $q_1$ to mode $q_2$. Using \eqref{eq:saltation}, we now derive $\mathbf{\Xi}$ for \eqref{HA} when the inverter switches from GFL to GFM. The Jacobian of the reset map in \eqref{HA} is given by
\vspace{-.15cm}
\begin{align}
\mathbf{D_x\mathrm{R}}(\mathbf{x} ^-)
=
\left[\begin{array}{cc}
\frac{\partial\mathrm{R_{gfl}}}{\mathbf{x}_{\mathrm{gfl}}} & \frac{\partial\mathrm{R_{gfl}}}{\mathbf{x}_{\mathrm{gfm}}} \\
\frac{\partial\mathrm{R_{gfm}}}{\mathbf{x}_{\mathrm{gfl}}} &
\frac{\partial\mathrm{R_{gfm}}}{\mathbf{x}_{\mathrm{gfm}}}
\end{array}\right]=
\begin{bmatrix}
\mathbf{I}_6 & \mathbf{0}_{6\times7}\\[3pt]
\mathbf{0}_{7\times6} & \mathbf{S}_{7\times7}
\end{bmatrix}
\end{align}

\vspace{-.15cm}
\noindent
$\mathbf{I}$ (resp., $\mathbf{0}$) is an identity (resp., a zero) matrix of appropriate dimensions, 
$\mathbf{S}=\mathbf{I}_7 - \mathbf{e}_{\theta_{\mathrm{gfm}}} \mathbf{e}_{\theta_{\mathrm{gfm}}}^{\top}=\mathrm{diag}(1,1,1,1,1,1,0)$, and the canonical basis vector $\mathbf{e}_{\theta_{\mathrm{gfm}}} =[0\; 0\; 0\; 0\; 0\; 0\; 1]^{\top}$. For a voltage guard $g_v(\mathbf{x}) = |v_{\text{grid}}| - v_{\text{th}}$ with $|v^{\mathrm{grid}}| = \sqrt{v_d^2 + v_q^2}$,
\vspace{-.15cm}
\begin{align}
\nabla g_v(\mathbf{x}^-) =
\big[\, \mathbf{0}_6\quad
\tfrac{v_d^-}{|v^{\mathrm{grid}}|^-}\quad
\tfrac{v_q^-}{|v^{\mathrm{grid}}|^-}\quad
\mathbf{0}_{5}\,\big]^{\top}
\end{align}

\vspace{-.15cm}
For a frequency guard $g_\omega(\mathbf{x}) = |\Delta\omega| - \omega_{\mathrm{th}}$ with
$\Delta\omega = \omega(\mathbf{x}_{\mathrm{gfl}}) - \omega_0$,
\vspace{-.25cm}
\begin{align}
\nabla g_\omega(\mathbf{x}^-)
=
\bigg[
\frac{\partial \omega}{\partial x_{\mathrm{gfl}}}
\;\; \mathbf{0}_7
\bigg]^{\top}
\end{align}

\vspace{-.15cm}
The resulting saltation matrix is of the form:
\vspace{-.15cm}
\begin{align}
\mathbf{\Xi} &= 
\underbrace{\big[\mathbf{I}_{13}
+ (\mathbf{e}_{\phi_{\mathrm{gfm}}} \mathbf{e}_{\phi_{\mathrm{pll}}}^{\top}
- \mathbf{e}_{\phi_{\mathrm{gfm}}} \mathbf{e}_{\phi_{\mathrm{gfm}}}^{\top})\big]}_{\mathbf{D}_{\mathbf{x}} \mathrm{R}_{q_{1} q_{2}}^{-} } \nonumber \\
&\quad + 
\frac{1}{\alpha}
\underbrace{
\begin{bmatrix}
-\,\mathbf{f}_{\mathrm{gfl}}(\mathbf{x}_{\mathrm{gfl}},\mathbf{u}_{\mathrm{gfl}})\\
\;\;\mathbf{f}_{\mathrm{gfm}}\!\big(\mathbf{x}_{\mathrm{gfm}},\mathbf{u}_{\mathrm{gfm}}(p_0^{+},q_0^{+}),r^{+},\ell^{+}\big)
\end{bmatrix}
}_{\mathbf{f}_{q_{2}}^{+} - \mathbf{D}_{\mathbf{x}} \mathrm{R}_{q_{1} q_{2}}^{-}} \nonumber \\[4pt]
&\quad \cdot
\underbrace{
\begin{bmatrix}
\nabla_{\mathbf{x}_{\mathrm{gfl}}} \mathbf{g}(\mathbf{x})^{\top} \quad 
\nabla_{\mathbf{x}_{\mathrm{gfm}}} \mathbf{g}(\mathbf{x})^{\top}
\end{bmatrix}
}_{\left(\nabla\mathbf{g}_{q_{1} q_{2}}^{-}\right)^{\top}}
\end{align}

\vspace{-.15cm}
\noindent
where \(\alpha \coloneqq \left(\nabla \mathbf{g}_{q_{1} q_{2}}^{-}\right)^{\top} \mathbf{f}_{q_{1}}^{-}\). 

Note that the main difference between the continuous \cite{Netto2016} and hybrid EKF formulations occurs when a guard condition is triggered; specifically, how, in that case, a state estimate and covariance are updated during a prediction or correction step. 

\noindent
\emph{Remark 1:} A \emph{continuous} formulation could, in theory, be used to approximate the switching logic with a smooth convex interpolation between GFL and GFM dynamics, as follows:
\vspace{-.15cm}
\begin{equation}
\mathbf{f}^{\mathrm{cont}}(\mathbf{x}) = \sigma(v^{\mathrm{grid}})\, \mathbf{f}_{\mathrm{gfl}}(\mathbf{x})
+ \big(1-\sigma(v^{\mathrm{grid}})\big)\, \mathbf{f}_{\mathrm{gfm}}(\mathbf{x})
\end{equation}

\vspace{-.15cm}
\noindent
where $\sigma(v^{\mathrm{grid}})=\big(1+\exp[-k(v^{\mathrm{grid}}-V_{\mathrm{th}})]\big)^{-1}$ 
is a logistic weighting function, and $k$ is a gain parameter that controls the sharpness of the transition. In the next section, we note that this approximation using a continuous formulation yields lower estimation accuracy than the hybrid formulation.\qed

\begin{table}[ht]
\centering
\caption{Parameters of the hybrid GFL--GFM inverter model}
\begin{tabular}{lll}
\hline
\textbf{Parameter} & \textbf{Description} & \textbf{Value} \\ \hline
{$r_f$} & Filter resistance & 1.89 pu \\
{$\ell_f$} & Filter inductance & 0.02 pu \\
{$r_g$} & Grid resistance (SCR = 5) & 0.02 pu \\
{$\ell_g$} & Grid inductance (SCR = 5) & 0.01 pu \\
$k_p^{\mathrm{pll}}, k_i^{\mathrm{pll}}$ & PLL proportional and integral gains & 0.02, 0.10 \\
$k_{pd}$ & PLL damping gain & 0.10 \\
$k_p^{i}, k_i^{i}$ & Current controller gains & 1.2, 40 \\
$m_p$ & Active power droop coefficient & 0.02 \\
$n_q$ & Reactive power droop coefficient & 0.012 \\
$v_0$ & Nominal voltage reference & 1.00 pu \\
$r_{vi}$, $\ell_{vi}$ & Virtual impedance & 0.05, 0.05 pu \\
$i_{\mathrm{th}}, i_{\max}$ & Current thresholds & 0.40, 1.20 pu \\
$v_{\mathrm{th}}$ & Voltage threshold for mode transition & 0.90 pu \\
$\Delta\omega_{\mathrm{th}}$ & Frequency deviation threshold & $2\pi \times 0.05$ rad/s \\
$\omega_0$ & Nominal angular frequency & $2\pi \times 30$ rad/s \\
$\mathbf{Q}_{q_k}$ & Process noise covariance & $10^{-6}\mathbf{I}_6$ \\
$\mathbf{\Sigma}_{q_k}$ & Measurement noise covariance & $\mathrm{diag}(10^{-5}\text{:}10^{-4})$ \\ \hline
\end{tabular}
\label{tab:params}
\end{table}

\begin{table*}[!t]
\centering
\scriptsize
\caption{RMSE comparison for extended Kalman filter state estimation using hybrid and continuous underlying models, for near-switch and overall periods}
\label{modeswitching}
\begin{tabular}{lcccc|cccc}
\hline
& \multicolumn{4}{c}{\textbf{Near-switch RMSE}} & \multicolumn{4}{c}{\textbf{Overall RMSE}} \\
\cline{2-5} \cline{6-9}
\textbf{Underlying model} 
& $i_d$ & $i_q$ & $v_d$ & $v_q$
& $i_d$ & $i_q$ & $v_d$ & $v_q$ \\
\hline
\rowcolor{gray!15}
Hybrid system model 
& 5.6$\times10^{-3}$ 
& 5.4$\times10^{-5}$ 
& 2.30$\times10^{-4}$ 
& 3.43$\times10^{-4}$
& 2.13$\times10^{-4}$ 
& 2.87$\times10^{-4}$ 
& 2.70$\times10^{-4}$ 
& 7.81$\times10^{-4}$ \\
Continuous-time model 
& 7.40$\times10^{-2}$ 
& 5.69$\times10^{-4}$ 
& 2.75$\times10^{-2}$ 
& 2.35$\times10^{-2}$
& 5.17$\times10^{-4}$ 
& 6.60$\times10^{-4}$ 
& 2.27$\times10^{-2}$ 
& 2.07$\times10^{-2}$ \\
\hline
\end{tabular}
\end{table*}

\begin{figure*}[ht]
\centering
\begin{subfigure}[b]{0.32\textwidth}
\centering
\includegraphics[width=\linewidth]{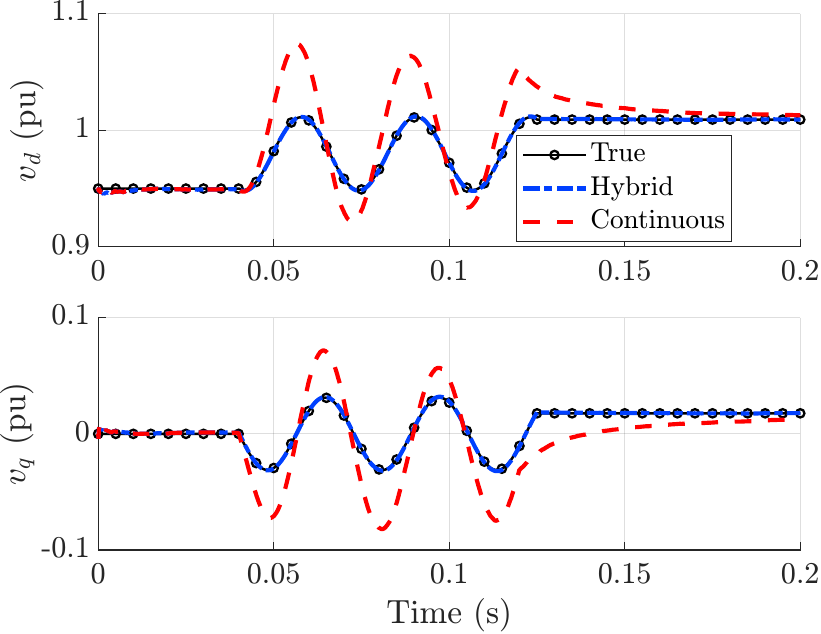}
\caption{Voltage variations.}
\label{fig:voltages}
\end{subfigure}
\hfill
\begin{subfigure}[b]{0.32\textwidth}
\centering
\includegraphics[width=\linewidth]{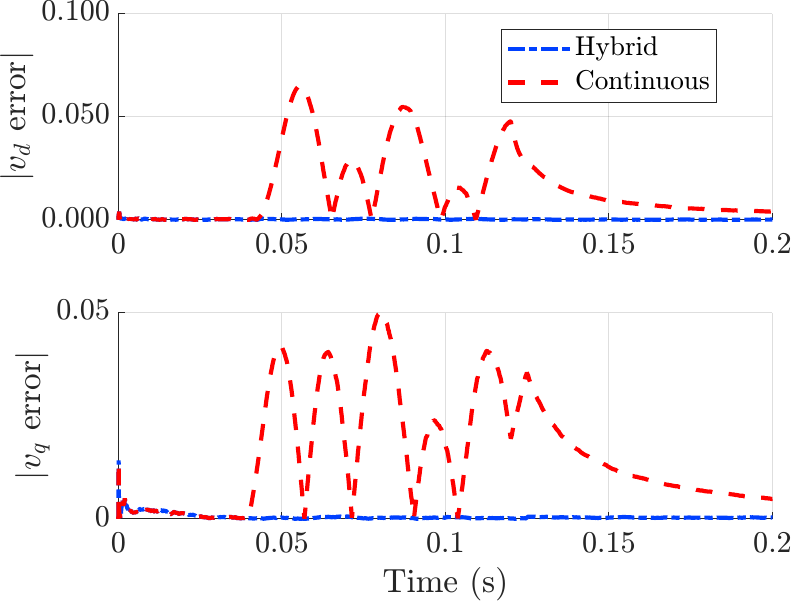}
\caption{Absolute voltage error.}
\label{fig:abs_error_voltages}
\end{subfigure}
\hfill
\begin{subfigure}[b]{0.32\textwidth}
\centering
\includegraphics[width=\linewidth]{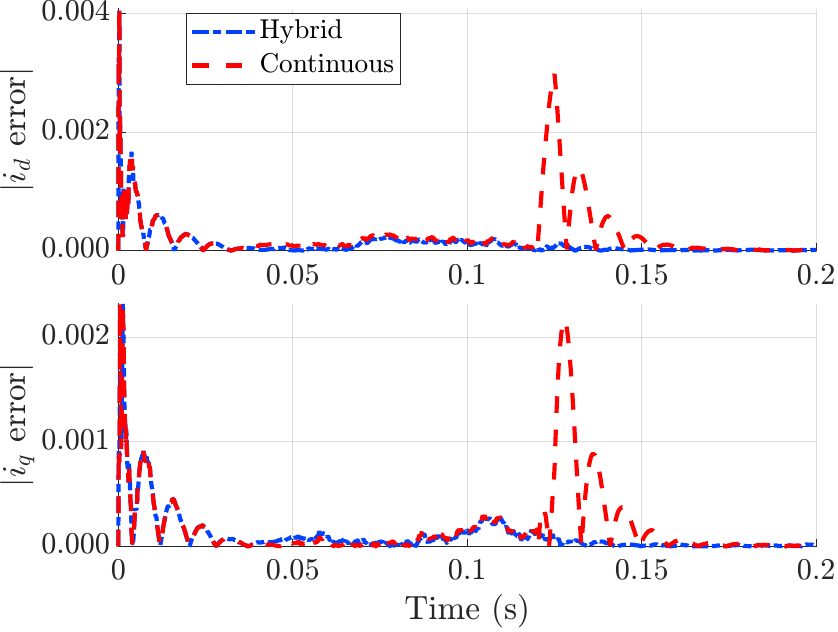}
\caption{Absolute current error.}
\label{fig:abs_error_currents}
\end{subfigure}
\caption{Comparison of voltage and current responses for hybrid and continuous models, showing voltage variations, voltage errors, and current errors.}
\label{fig:hybrid_three_comparison2}
\end{figure*}

\vspace{-.4cm}
\section{Numerical results}

The parameters for the results in this section are presented in Table~\ref{tab:params}. Figs.~\ref{fig:hybrid_voltage} and~\ref{fig:hybrid_current} show the voltage and current trajectories of the hybrid inverter model in~\eqref{HA}, including zoomed-in views around the switching instant. Each figure displays three trajectories: the system simulated under GFL and GFM dynamics only, and the hybrid trajectory from \eqref{HA}. When the guard conditions are satisfied, the voltage and current governed by the GFL dynamics only oscillates, while the hybrid system transitions to GFM control and maintains stable behavior. Fig.~\ref{fig:no_reset} shows the same transition without applying the reset map, where a noticeable surge appears in the current at the switching instant. This demonstrates that the reset map preserves stable trajectories during mode transitions and prevents inconsistent state initialization and overcurrent in the GFM mode.

Figs.~\ref{fig:voltages}--\ref{fig:abs_error_currents} present the EKF estimation results for the hybrid inverter model during transitions between GFL and GFM modes in both directions. These plots compare the hybrid EKF with an EKF based on the continuous approximation discussed in Remark 1. The smoothed model exhibits large transient deviations during switching events because it cannot faithfully represent switching actions. Table~II summarizes the root-mean-square error (RMSE) between estimated and true state values evaluated over the whole trajectory and in a near-switching window. The larger RMSE of the continuous model is consistent with its innovation statistics computed from \eqref{eq:measurementupdate}; that is, the normalized innovation squared, which tests the match between predicted covariance and observed innovations and remains within the 95\% $\chi^2_4$ band for about 95\% of samples under the hybrid EKF, but only 25--31\% for the continuous approximation. This inconsistency reflects the underestimation of uncertainty near mode transitions.
\vspace{-.3cm}
\section{Conclusion}
We developed a hybrid system framework for inverters that switch between GFL and GFM control schemes, with reset maps that maintain phase, frequency, and droop continuity during mode transitions. When integrated with an EKF, the hybrid model improved estimation accuracy and reduced bias near mode transitions compared to continuous, smoothed models. Future work will expand to multiple parallel inverters with large-scale switching events at the point of interconnection with the grid and will examine the reachability of the hybrid system.

\bibliographystyle{IEEEtran}
\bibliography{lib}

\end{document}

%% file: fig1.tex
\definecolor{ffqqqq}{rgb}{1,0,0}
\begin{tikzpicture}[line cap=round,line join=round,>=triangle 45,x=1cm,y=1cm]
\clip(0.30,-0.10) rectangle (11.80,6.20);
\draw (0.18,5.62) node[anchor=north west] {$+$};
\draw (3.45,5.10) node[anchor=north west] {$i_d\!+\!ji_q$};
\draw (4.05,6.05) node[anchor=north west] {$v_d\!+\!jv_q$};
\draw (6.85,6.10) node[anchor=north west] {$v^{\mathrm{filt}}_d\!+\!jv^{\mathrm{filt}}_q$};
\draw (9.70,6.15) node[anchor=north west] {$v^{\mathrm{grid}}_d\!+\!jv^{\mathrm{grid}}_q$};
\draw (5.35,5.70) node[anchor=north west] {$r_f\!+\!j\ell_f$};
\draw (8.35,5.70) node[anchor=north west] {$r_g\!+\!j\ell_g$};
\draw [line width=0.4pt,dash pattern=on 1pt off 1pt] (5.0,5.20)-- (5.0,5.6);
\draw [line width=0.4pt,dash pattern=on 1pt off 1pt] (8.0,5.20)-- (8.0,5.6);
\draw [line width=0.4pt,dash pattern=on 1pt off 1pt] (10.8,5.20)-- (10.8,5.6);
\draw [color=black] (10.8,5.1) circle (1.5pt);
\draw [color=black] (8.0,5.1) circle (1.5pt);
\draw [color=black] (5.0,5.1) circle (1.5pt);
\draw (4.264311289534664,4.45) node[anchor=north west] {\parbox{1.641029860715749 cm}{Inner\\loop}};
\draw (0.60,1.95) node[anchor=north west] {PLL};
\draw (3.05,1.95) node[anchor=north west] {\parbox{1.641029860715749 cm}{Outer\\loop}};
\draw (5.05,1.95) node[anchor=north west] {\parbox{1.8974418050020485 cm}{Droop\\control}};
\draw (7.55,1.95) node[anchor=north west] {\parbox{1.8974418050020485 cm}{Voltage\\control}};
\draw [line width=0.8pt,color=ffqqqq] (9.3,2.9)-- (9.8,2.9);
\draw (9.75,3.15) node[anchor=north west] {\parbox{3.3845329888007316 cm}{Control \&\\commun.}};
\draw (10.30,4.00) node[anchor=north west] {Eq. (1)};
\draw (1.85,1.00) node[anchor=north west] {Eq. (2)};
\draw (6.35,1.00) node[anchor=north west] {Eq. (3)};
\draw (4.20,0.45) node[anchor=north west] {$p^{\mathrm{ref}},q^{\mathrm{ref}}$};
\draw (0.40,2.75) node[anchor=north west] {$\omega, \theta^{\mathrm{pll}}$};
\draw (5.50,2.70) node[anchor=north west] {$\omega, \theta$};
\draw (3.35,2.75) node[anchor=north west] {$i^{\mathrm{ref}}$};
\draw (7.95,2.75) node[anchor=north west] {$i^{\mathrm{ref}}$};
\fill[line width=0pt,fill=black,fill opacity=1] (1.94,5.22) -- (1.94,4.98) -- (1.7,5.1) -- cycle;
\fill[line width=0pt,fill=black,fill opacity=0.05] (3.5,6.2) -- (3.5,3.4) -- (11.7,3.4) -- (11.7,6.2) -- cycle;
\fill[line width=0pt,fill=black,fill opacity=0.05] (0.4,2.2) -- (0.4,0.5) -- (4.6,0.5) -- (4.6,2.2) -- cycle;
\fill[line width=0pt,fill=red,fill opacity=0.1] (4.9,0.5) -- (4.9,2.2) -- (9.1,2.2) -- (9.1,0.5) -- cycle;
\draw [line width=0.4pt] (2.9,4.2)-- (1.1,4.2);
\draw [line width=0.4pt] (1.1,4.2)-- (1.1,6);
\draw [line width=0.4pt] (1.1,6)-- (2.9,6);
\draw [line width=0.4pt] (2.9,6)-- (2.9,4.2);
\draw [line width=0.4pt] (2,5.1) circle (0.8cm);
\draw [line width=0.8pt] (2,5.4)-- (2,5.95);
\draw [line width=0.8pt] (2,5.1)-- (2,4.25);
\draw [line width=0.8pt] (1.6,5.1)-- (1.15,5.1);
\draw [line width=0.8pt] (1.7,4.8)-- (2,4.8);
\draw [line width=0.8pt] (2,5.4)-- (1.7,5.4);
\draw [line width=0.8pt] (1.6,5.52)-- (1.6,4.68);
\draw [line width=0.8pt] (2,5.7)-- (2.4,5.7);
\draw [line width=0.8pt] (2,4.5)-- (2.4,4.5);
\draw [line width=0.8pt] (1.7,5.52)-- (1.7,5.28);
\draw [line width=0.8pt] (1.7,5.22)-- (1.7,4.98);
\draw [line width=0.8pt] (1.7,4.92)-- (1.7,4.68);
\draw [line width=0.8pt] (2.4,4.5)-- (2.4,4.9);
\draw [line width=0.8pt] (2.22,4.9)-- (2.58,4.9);
\draw [line width=0.8pt] (2.58,4.9)-- (2.4,5.26);
\draw [line width=0.8pt] (2.4,5.26)-- (2.22,4.9);
\draw [line width=0.8pt] (2.22,5.26)-- (2.58,5.26);
\draw [line width=0.8pt] (2.4,5.7)-- (2.4,5.26);
\draw [line width=0.8pt] (2,5.1)-- (1.94,5.1);
\draw [line width=0.4pt] (0.6,5.8)-- (0.6,5.2);
\draw [line width=0.4pt] (0.6,4.4)-- (0.6,5);
\draw [line width=0.4pt] (0.4,5.2)-- (0.8,5.2);
\draw [line width=0.4pt] (0.4,5)-- (0.8,5);
\draw [line width=0.4pt] (0.6,5.8)-- (1.1,5.8);
\draw [line width=0.4pt] (0.6,4.4)-- (1.1,4.4);
\draw [line width=0.4pt] (5.5,5.1)-- (2.9,5.1);
\draw [line width=0.4pt] (5.5,5.2)-- (5.5,5);
\draw [line width=0.4pt] (5.5,5.2)-- (6.5,5.2);
\draw [line width=0.4pt] (6.5,5.2)-- (6.5,5);
\draw [line width=0.4pt] (6.5,5)-- (5.5,5);
\draw [line width=0.4pt] (6.5,5.1)-- (8.5,5.1);
\draw [line width=0.4pt] (8.5,5.2)-- (8.5,5);
\draw [line width=0.4pt] (8.5,5)-- (9.5,5);
\draw [line width=0.4pt] (9.5,5)-- (9.5,5.2);
\draw [line width=0.4pt] (9.5,5.2)-- (8.5,5.2);
\draw [line width=0.4pt] (9.5,5.1)-- (11,5.1);
\draw [line width=0.8pt] (2.98,4.96)-- (3.14,5.24);
\draw [line width=0.8pt] (3.08,4.96)-- (3.24,5.24);
\draw [line width=0.8pt] (3.18,4.96)-- (3.34,5.24);
\draw [line width=0.8pt] (4.3,5.1)-- (3.7,5.1);
\draw [line width=0.8pt,color=ffqqqq] (2.6,3.8)-- (1.6,3.8);
\draw [line width=0.8pt,color=ffqqqq] (1.6,3.8)-- (1.6,4.12);
\draw [line width=0.8pt,color=ffqqqq] (1.8,4.12)-- (1.8,3.8);
\draw [line width=0.8pt,color=ffqqqq] (2,4.12)-- (2,3.8);
\draw [line width=0.8pt,color=ffqqqq] (2.2,4.12)-- (2.2,3.8);
\draw [line width=0.8pt,color=ffqqqq] (2.4,4.12)-- (2.4,3.8);
\draw [line width=0.8pt,color=ffqqqq] (2.6,4.12)-- (2.6,3.8);
\draw [line width=0.8pt,color=ffqqqq] (2.6,3.8)-- (4,3.8);
\draw [line width=0.4pt] (4,4.5)-- (4,3.5);
\draw [line width=0.4pt] (4,4.5)-- (5.5,4.5);
\draw [line width=0.4pt] (5.5,4.5)-- (5.5,3.5);
\draw [line width=0.4pt] (5.5,3.5)-- (4,3.5);
\draw [line width=0.8pt,color=ffqqqq] (4.75,3.05)-- (4.75,3.498);
\draw [line width=0.8pt,color=ffqqqq] (4.55,2.7)-- (1.5,2.7);
\draw [line width=0.8pt,color=ffqqqq] (4.95,2.7)-- (8,2.7);
\draw [line width=0.4pt] (4.5,2)-- (4.5,1);
\draw [line width=0.4pt] (4.5,1)-- (3,1);
\draw [line width=0.4pt] (3,1)-- (3,2);
\draw [line width=0.4pt] (3,2)-- (4.5,2);
\draw [line width=0.4pt] (2,2)-- (2,1);
\draw [line width=0.4pt] (2,1)-- (0.5,1);
\draw [line width=0.4pt] (0.5,1)-- (0.5,2);
\draw [line width=0.4pt] (0.5,2)-- (2,2);
\draw [line width=0.4pt] (5,2)-- (5,1);
\draw [line width=0.4pt] (5,1)-- (6.5,1);
\draw [line width=0.4pt] (6.5,1)-- (6.5,2);
\draw [line width=0.4pt] (6.5,2)-- (5,2);
\draw [line width=0.4pt] (7.5,2)-- (7.5,1);
\draw [line width=0.4pt] (7.5,1)-- (9,1);
\draw [line width=0.4pt] (9,1)-- (9,2);
\draw [line width=0.4pt] (9,2)-- (7.5,2);
\draw [line width=0.8pt,color=ffqqqq] (1.5,2.7)-- (1.5,2);
\draw [line width=0.8pt,color=ffqqqq] (8,2.7)-- (8,2);
\draw [line width=0.8pt,color=ffqqqq] (5.5,2)-- (5.5,2.7);
\draw [line width=0.8pt,color=ffqqqq] (4,2)-- (4,2.7);
\draw [line width=0.8pt,color=ffqqqq] (2,1.5)-- (3,1.5);
\draw [line width=0.8pt,color=ffqqqq] (6.5,1.5)-- (7.5,1.5);
\draw [line width=0.8pt,color=ffqqqq] (4,0.7)-- (4,0.95);
\draw [line width=0.8pt,color=ffqqqq] (5.5,0.7)-- (5.5,0.95);
\draw [line width=0.8pt,color=ffqqqq] (5.5,0.7)-- (4,0.7);
\draw [line width=0.8pt,color=ffqqqq] (4.75,0.7)-- (4.75,0.4);
\draw [line width=1.6pt] (4.75,3.05)-- (4.568,2.682);
\draw [line width=0.4pt] (1.6,3.2)-- (1.7,3.2);
\draw [line width=0.4pt] (1.7,3.2)-- (1.7,3.7);
\draw [line width=0.4pt] (1.7,3.7)-- (1.8,3.7);
\draw [line width=0.4pt] (1.8,3.7)-- (1.8,3.2);
\draw [line width=0.4pt] (1.8,3.2)-- (1.9,3.2);
\draw [line width=0.4pt] (1.9,3.2)-- (1.9,3.7);
\draw [line width=0.4pt] (1.9,3.7)-- (2,3.7);
\draw [line width=0.4pt] (2,3.7)-- (2,3.2);
\draw [line width=0.4pt] (2,3.2)-- (2.1,3.2);
\draw [line width=0.4pt] (2.1,3.2)-- (2.1,3.7);
\draw [line width=0.4pt] (2.1,3.7)-- (2.2,3.7);
\draw [line width=0.4pt] (2.2,3.7)-- (2.2,3.2);
\draw [line width=0.4pt] (2.2,3.2)-- (2.3,3.2);
\draw [line width=0.4pt] (2.3,3.2)-- (2.3,3.7);
\draw [line width=0.4pt] (2.3,3.7)-- (2.4,3.7);
\draw [line width=0.4pt] (2.4,3.7)-- (2.4,3.2);
\draw [line width=0.4pt] (2.4,3.2)-- (2.5,3.2);
\draw [line width=0.4pt] (2.5,3.2)-- (2.5,3.7);
\draw [line width=0.4pt] (2.5,3.7)-- (2.6,3.7);
\draw [line width=0.4pt] (2.6,3.7)-- (2.6,3.2);
\draw [line width=0.4pt] (2.6,3.2)-- (2.7,3.2);
%
\begin{scope}[shift={(1.01,2.68)}, scale=0.22, transform shape]
\draw[line width=0.4pt] (3,3.5) sin (4,4.5);
\draw[line width=0.4pt] (4,4.5) cos (5,3.5);
\draw[line width=0.4pt] (5,3.5) sin (6,2.5);
\draw[line width=0.4pt] (6,2.5) cos (7,3.5);
\end{scope}
\begin{scriptsize}
\draw [fill=black] (2,5.7) circle (1pt);
\draw [fill=black] (2,4.5) circle (1pt);
\draw [fill=black,shift={(4.3,5.1)},rotate=270] (0,0) ++(0 pt,1.5pt) -- ++(1.299038105676658pt,-2.25pt)--++(-2.598076211353316pt,0 pt) -- ++(1.299038105676658pt,2.25pt);
\draw [fill=ffqqqq,shift={(1.6,4.12)}] (0,0) ++(0 pt,2.25pt) -- ++(1.9485571585149868pt,-3.375pt)--++(-3.8971143170299736pt,0 pt) -- ++(1.9485571585149868pt,3.375pt);
\draw [fill=ffqqqq,shift={(1.8,4.12)}] (0,0) ++(0 pt,2.25pt) -- ++(1.9485571585149868pt,-3.375pt)--++(-3.8971143170299736pt,0 pt) -- ++(1.9485571585149868pt,3.375pt);
\draw [fill=ffqqqq,shift={(2,4.12)}] (0,0) ++(0 pt,2.25pt) -- ++(1.9485571585149868pt,-3.375pt)--++(-3.8971143170299736pt,0 pt) -- ++(1.9485571585149868pt,3.375pt);
\draw [fill=ffqqqq,shift={(2.2,4.12)}] (0,0) ++(0 pt,2.25pt) -- ++(1.9485571585149868pt,-3.375pt)--++(-3.8971143170299736pt,0 pt) -- ++(1.9485571585149868pt,3.375pt);
\draw [fill=ffqqqq,shift={(2.4,4.12)}] (0,0) ++(0 pt,2.25pt) -- ++(1.9485571585149868pt,-3.375pt)--++(-3.8971143170299736pt,0 pt) -- ++(1.9485571585149868pt,3.375pt);
\draw [fill=ffqqqq,shift={(2.6,4.12)}] (0,0) ++(0 pt,2.25pt) -- ++(1.9485571585149868pt,-3.375pt)--++(-3.8971143170299736pt,0 pt) -- ++(1.9485571585149868pt,3.375pt);
\draw [color=black] (4.55,2.7) circle (2pt);
\draw [color=black] (4.95,2.7) circle (2pt);
\draw [color=black] (4.75,3.05) circle (2pt);
\draw [fill=ffqqqq,shift={(4,0.95)}] (0,0) ++(0 pt,1.5pt) -- ++(1.299038105676658pt,-2.25pt)--++(-2.598076211353316pt,0 pt) -- ++(1.299038105676658pt,2.25pt);
\draw [fill=ffqqqq,shift={(5.5,0.95)}] (0,0) ++(0 pt,1.5pt) -- ++(1.299038105676658pt,-2.25pt)--++(-2.598076211353316pt,0 pt) -- ++(1.299038105676658pt,2.25pt);
\end{scriptsize}
\end{tikzpicture}